\documentclass{elsart4}

\usepackage{amssymb}

\usepackage{graphics}
\usepackage{graphicx}
\usepackage{epsfig}
\usepackage{amssymb}
\journal{Physica B}

\begin{document}

\begin{frontmatter}


 \title{Photoconduction in CDW conductors} 
 \author{S.V. Zaitsev-Zotov\corauthref{cor1},} 
 \author{V.E. Minakova, V.F.  Nasretdinova, and S.G. Zybtsev} 
 \address{Kotel'nikov IRE RAS, Mokhovaya 11, bld.7, 125009 Moscow, Russia} 
 \corauth[cor1]{ Tel: +7 (495) 629 33 94; FAX:+7 (495) 629 36 78; e-mail: serzz@cplire.ru}





\begin{abstract}
Photoconduction study of quasi-1D conductors allows to distinguish between the single-particle and collective {\it linear} conduction, investigate the effect of screening on collective transport and obtain interesting new details of the electronic energy structure of pure and doped CDW conductors. Here we present results of photoconduction study in quasi-1D conductors  {\it o}-TaS$_3$, K$_{0.3}$MoO$_3$, and NbS$_3$(I).
\end{abstract}

\end{frontmatter}

\section{Introduction}

Effect of illumination on transport properties of the quasi-one-dimensional (q-1D) conductors with the charge-density waves (CDW) was firstly observed by Ogawa and co-authors 10 years ago in  K$_{0.3}$MoO$_3$ \cite{ogawa}. Since then similar behavior was observed in orthorhombic TaS$_3$ \cite{zzminajetpl,zzminaprl}, and a number of questions raised by Ogawa and co-authors were answered and reinterpreted in terms of photoconduction \cite{zzminajetpl,zzecrys2005,zzminaprl}, rather than direct influence of light on the CDW \cite{ogawa}. In addition, photoconduction was successfully employed for energy-structure study of {\it o}-TaS$_3$, K$_{0.3}$MoO$_3$ and NbS$_3$ (phase I). As a result, it gave Peierls gap values, demonstrated a relatively sharp Peierls gap edge (no pseudo-gap behavior) and even revealed intra-gap states \cite{naszzjetpl}. Here we present a brief review of the main features of photoconduction in quasi-1D conductors {\it o}-TaS$_3$, K$_{0.3}$MoO$_3$ and NbS$_3$(I).

{\it o}-TaS$_3$ and  K$_{0.3}$MoO$_3$ are typical CDW conductors with fully gaped Fermi surfaces and the Peierls transitions at $T_P=220$ K and 180 K respectively \cite{review}. NbS$_3$(I) is a quasi-1D conductor with a chain-like structure and two-fold commensurable superstructure along the chains \cite{NbS3}. It demonstrates non-metallic behavior over entire temperature range  with the low-temperature activation energy $\sim 4000$~K ($T\lesssim 200$ K) \cite{NbS3rt}. No metal-insulator transition is observed in this compound.

\section{Experimental}

We have studied nominally pure {\it o}-TaS$_3$ crystals ($E_T(120{\rm\ K}) \lesssim 1$~V/cm) provided by R.E. Thorne (Cornell University), and F. Levy (Inst\'{i}tute de Phys\'{i}que Appliqu\'{e}e, Lausanne). Nb-doped crystals were provided by F. Levy, K$_{0.3}$MoO$_3$ crystals were provided by R.E. Thorne, Ti-doped {\it o}-TaS$_3$ and NbS$_3$(I) crystals were grown in Kotel'nikov IRE RAS. $E_T(120{\rm\ K}) \gg 10$~V/cm for all doped {\it o} -TaS$_3$ samples.

In {\it o}-TaS$_3$ the sample thickness was comparable or even smaller the light penetration depth ($t\sim 0.1 - 1\;\mu$m). Thinnest available samples of K$_{0.3}$MoO$_3$ and NbS$_3$(I) were chosen for study ($t\sim 10\;\mu$m).  Using thin samples allows one to reduce shunting of photoconduction by the bulk conduction and suppress or even eliminate the heating effect due to exceptionally good thermal contact with sapphire substrate.  Current terminals were made by indium cold soldering with a predeposited gold layer in the case of  K$_{0.3}$MoO$_3$. All the measurements were performed in the two-terminal configuration in the voltage-controlled regime. 

IR LED driven by meander-modulated current (modulation frequency $f= 4.5$ Hz) producing light flux $W=(10^{-6}-10^{-2}$) W/cm$^2$ at the sample position and with a wavelength of $\lambda = 0.94\;\mu$m was used for investigation of temperature- and intensity-dependent photoconductivity. As it will be clear from below, the photon energy is higher than the optical gap value in all studied materials. 

Photoconduction spectra were measured using a grating monochromator with a globar  ({\it o}-TaS$_3$, K$_{0.3}$MoO$_3$) or quartz lamp (NbS$_3$(I)) as a light source. The light intensity was modulated at a frequency 3.125 or 6.25 Hz by a light chopper, a lock in amplifier was used for photocurrent measurements. To reduce the effect of light absorption by the air, the monochromator was evacuated to a pressure below 1 Torr. Other details of the measurements can be found in our earlier publications \cite{zzminajetpl,zzecrys2005,zzminaprl,naszzjetpl}.

\section{Main features of photoconduction in {\it o}-TaS$_3$,  K$_{0.3}$MoO$_3$ and NbS$_3$(I)}

Photoconduction has a pronounced temperature dependence restricting the temperature region of its observation in all studied materials. In {\it o}-TaS$_3$ it can be observed only at $T\lesssim 100$~K, i.e. well below the Peierls transition temperature $T_P=220$~K \cite{zzminajetpl,zzecrys2005,zzminaprl}. In K$_{0.3}$MoO$_3$ the temperature region is even more restricted: a noticeable photocurrent signal appears only at temperatures below 25-30 K despite the comparable $T_P$ value to {\it o}-TaS$_3$. In NbS$_3$(I) photoconduction can be observed at temperatures below 220 K. Such a behavior is a consequence of the activation temperature dependence of nonequilibrium current relaxation time and the linear-to-quadratic recombination crossover \cite{zzminaprl}. Analysis of this crossover provides a way to distinguish between collective and single-particle linear conduction in {\it o}-TaS$_3$ (see below).

\begin{figure}
\includegraphics[width=7cm]{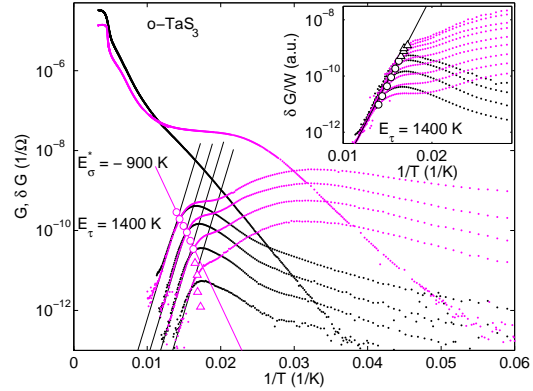}
\caption{Temperature variation of the linear conductance and photoconductance in two nominally pure {\it o}-TaS$_3$ crystals. Light modulation frequency $f=4.5$~Hz, $W= 0.01$, 0.1, 1, and 10 mW/cm$^2$. Inset shows photoconduction curves normalized by the light intensity (see text for details).}
\label{rtTaS3}
\end{figure}

\begin{figure}
\includegraphics[width=7cm]{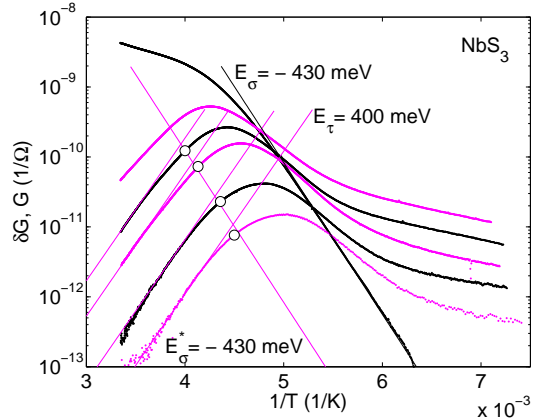}
\caption{Temperature variation of the linear conductance and photoconductance in NbS$_3$(I) at light intensities between 10 $\mu$W/cm$^2$ (lower curve) and 1 mW/cm$^2$ (upper curve).} \label{rtNbS3}
\end{figure}

Figs. \ref{rtTaS3} (black curves) and \ref{rtNbS3} show typical temperature variations of the dark linear conductance, $G(T)$, together with a set of photoconductance curves, $\delta G \equiv G(T,W)-G(T,0)$, measured at different light intensities, $W$, in {\it o}-TaS$_3$ and NbS$_3$(I) respectively. Though $G(T)$ curves are very different, $\delta G(T,W)$ curves have very similar shape and consist of high-temperature activation growth followed by maximum and further smooth decrease with a change of slope at $\sim 0.1$ of the value of $\delta G(T,W)$ in maximum.
The inset in Fig.\ref{rtTaS3} shows $\delta G/W$ curves. Collapsing of all the activation parts of $\delta G(T,W)$ into a single activation dependence proves realization of the linear recombination in the activation parts.
Similarly, plotting $\delta G/W^{1/2}$ proves existence the quadratic recombination region at temperatures below the temperature where $\delta G(T,W)$ reaches its maximum (see also \cite{zzecrys2005,zzminaprl}). 

Red curves in Figs. \ref{rtTaS3} show another type of behavior. 
Here the activation part of $\delta G(T,W)$ is the same, whereas the low-temperature parts of both $G(T)$ and $\delta G(T,W)$ has  structures which can be considered as indications of an additional low-temperature  contribution. It is interesting to note that the low-temperature part of $G(T)$ at $T\lesssim 80$ K resemble $G(T,W)$-dependence in NbS$_3$(I) (Fig. \ref{rtNbS3}).

The possibility to observe easily the quadratic recombination region is a feature of quasi-1D conductors: in usual semiconductors it can only be observed in very clean materials. This indicates absence of any other competitive recombination channels in quasi-1D conductors.

The central point for further analysis is the dependence of the temperature variation of the 
recombination time for linear recombination regime, $\tau_0$, on a position of the chemical potential, $\zeta$: 
\begin{equation} 
\tau_0\propto \frac{n_0p_0}{n_0+p_0}e^\frac{2\Delta}{kT}\propto e^{(\Delta-|\zeta|)/kT},
\label{eqacten}
\end{equation}
where $n_0\propto e^{-(\Delta-\zeta)/kT}$ and $p_0\propto e^{-(\Delta+\zeta)/kT}$ are dark concentrations of electrons and holes \cite{moss}. In accordance with Equation \ref{eqacten}, the activation energy $E_\tau=\Delta-|\zeta|$ depends on position of the chemical potential level and reaches its maximum value $\Delta$ at $\zeta=0$.  
This means that the optical gap in {\it o}-TaS$_3$ is noticeably bigger the transport gap value estimated from the linear conduction activation energy $E_\sigma\approx 800$~K. In NbS$_3$(I) both energies are practically the same. 

Another essential point is the analysis of the linear-to-quadratic recombination crossover \cite{zzminaprl} which let us possible to estimate the concentration of carriers responsible for photoconduction. Details of this analysis can be found in Ref. \cite{zzminaprl}. The main conclusion is that $\delta G$ at $\Delta n + \Delta p=n_0+p_0$ corresponds to 1.5 times smaller value than it would be expected in the linear recombination case. In the experiment this corresponds to the downward deviation from the activation law $\tau_0\propto e^{E_\tau/kT}$ by factor 1.5 (circles and triangles in the inset in Fig. \ref{rtTaS3}). It should be also taken into account that such a deviation appears also when $\tau_0 \gtrsim 1/f$, where $f$ is the light modulation frequency (marked by triangles in the inset). It lets us possible to estimate the temperature where $\tau_0 = 2\pi f$, as well as where $\Delta n + \Delta p=n_0+p_0$ (shown by circles in Figs. \ref{rtTaS3} and \ref{rtNbS3}). The activation energies of the respective conductance $E_\sigma^* \approx 800 - 1200$ K for {\it o}-TaS$_3$ and 0.43 eV for NbS$_3$(I).

\begin{figure}
\includegraphics[width=7cm]{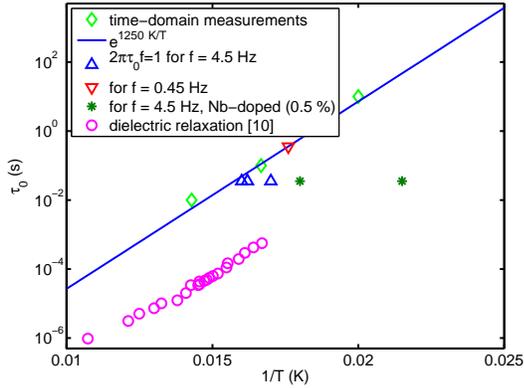}
\caption{Temperature variation of the relaxation time estimated by various methods in pure and Nb-doped {\em o}-TaS$_3$ samples together with the dielectric relaxation time for $\alpha$-process \protect\cite{starepsilon}.} 
\label{tau}
\end{figure}

Fig. \ref{tau} shows the activation dependence with the activation energy 1250 K \cite{zzminaprl} adjusted to fit the values of recombination time estimated from the time-domain study for the same sample (0.01 s, 1 s and 10 s at temperatures 70, 60 and 50 K respectively). It would be not very surprising, that $\tau(T)$ is practically the same as $\omega(T)$ for $\alpha$-process in the dielectric relaxation study \cite{starepsilon}.

In the case of NbS$_3$(I) the activation energies $E_\tau$, $E_\sigma$ and $E_\sigma^*$ are practically the same. This allows to conclude that photoconduction and the linear conduction are provided by the same current carriers. In {\it o}-TaS$_3$ the situation is dramatically different: $E_\tau$ is substantially bigger the transport activation energy over entire temperature range and especially at $T \lesssim 100$ K where $E_\sigma \approx 400$ K or even smaller. Note also that in both {\it o}-TaS$_3$ samples  $E_\sigma^*$ are practically the same despite substantially different $G(T)$ dependences (Fig. \ref{rtTaS3}). This proves that in {\it o}-TaS$_3$ the low-temperature linear conduction is not provided by a single particle mechanism, in agreement with the earlier analysis of low-temperature conduction anisotropy \cite{sambongi}.

\begin{figure}
\includegraphics[width=7cm]{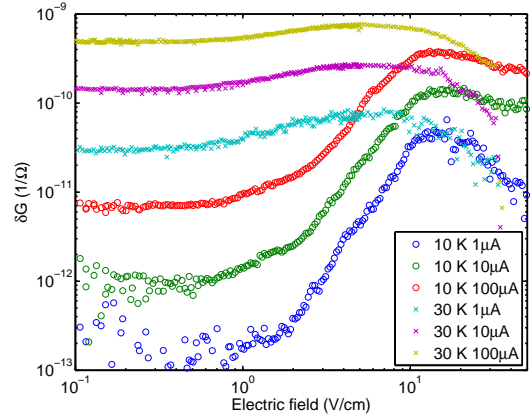}
\caption{Low-temperature development of electric-field dependence of photoconduction in {\it o}-TaS$_3$ sample with a shoulder on $G(T)$ (red curves in Fig. \protect\ref{rtTaS3}). Here LED current 1 mA corresponds approximately to 0.1 mW/cm$^2$ on a sample position.} 
\label{anomaly}
\end{figure}

Electric-field effect on photoconduction is different in samples with the activation and ``shoulder-like'' G(T) (respectively black and red curves in Fig. \ref{rtTaS3}):  application of the electric field enhances initially $\delta G$ in the sample with the shoulder on $G(T)$ (Fig. \ref{anomaly}).

\section{Effect of illumination on CDW transport and metastable states}

\begin{figure}
\includegraphics[width=6.5cm]{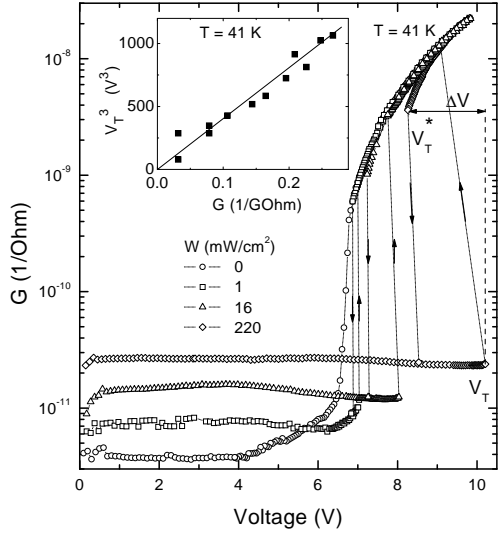}
\caption{Voltage-dependent conduction of a {\it o}-TaS$_3$ sample as a function of illumination intensity. Data of Ref. \cite{zzminajetpl}.} 
\label{ivs}
\end{figure}

Effect of steady illumination on the CDW threshold behavior in K$_{0.3}$MoO$_3$ was the first observation of light-induced changes in kinetics of the CDW \cite{ogawa}. This effect was reproduced and studied in details in {\it o}-TaS$_3$ \cite{zzminajetpl,zzecrys2005}. Fig. \ref{ivs} shows a typical set of I-V curves measured at different light intensities. It was found, in particular, that at a fixed temperature the threshold field for onset of CDW sliding, $E_T$, depends on conduction as $E_T\propto G^{1/3}$ (inset in Fig. \ref{ivs}).  This behavior results from changes of screening of CDW deformations by nonequilibrium carriers and corresponds to the case of 1D pinning \cite{zzminajetpl,zzecrys2005}.

Other interesting features are appearance of hysteresis in I-V curves and dissapearance of photoconduction at $E\gg E_T$ (Fig. \ref{ivs}). The later contradicts to the well-known scaling between nonlinear-and linear conduction \cite{ong}. 

\begin{figure}
\includegraphics[width=6.5cm]{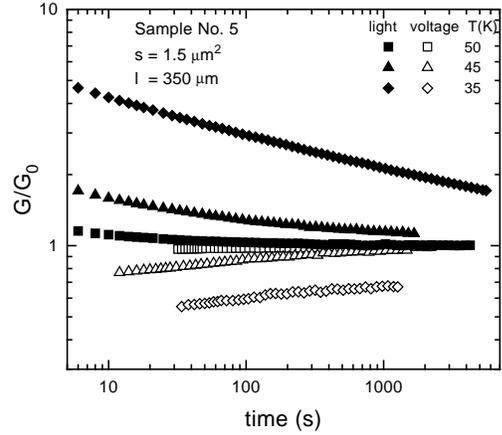}
\caption{Relaxation of metastable states produced by application of the voltage pulse (open symbols) and a light pulse (close symbols) in {\it o}-TaS$_3$ sample. All the curves are normalized by the ``equilibrium conduction'' $G_0$ obtained after cooling back from $T=120$~K to remove ``prehistory'' effect.  Data of Ref. \protect\cite{zzecrys2005}.} 
\label{relax}
\end{figure}

Fig. \ref{relax} shows relaxation of the linear conduction ($V<V_T$) after application of high voltage ($V>V_T$) and light pulses. Similarity between both relaxation processes let us to conclude that the physical mechanism of relaxation is also similar. Note that the temperature-induced metastability well-known for CDW conductors \cite{review} practically disappears below 50 K {\it o}-TaS$_3$ and should not be confused with the observed phenomena. Similarity of time scales for processes presented on Figs. \ref{tau} and  \ref{relax} leads us to the simplest conclusion on single-electron origin of this long-term relaxation.

\section{Photoconduction energy spectra}

\begin{figure}[b]
\includegraphics[width=6.5cm]{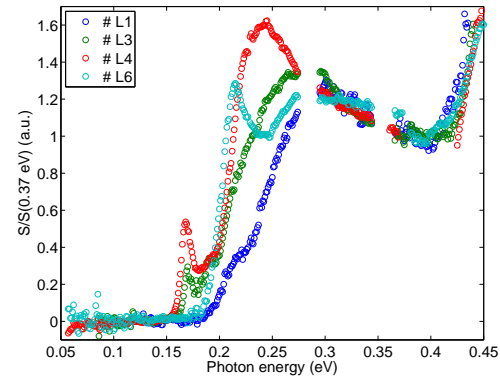}
\caption{Spectral dependencies of normalized photoconduction in four {\it o}-TaS$_3$ samples. Dara of Ref. \protect\cite{naszzjetpl}} 
\label{spTaS$_3$}
\end{figure}

\begin{figure}
\includegraphics[width=6.5cm]{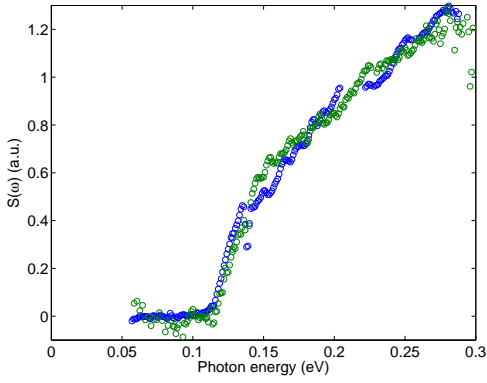}
\caption{Spectral dependencies of normalized photoconduction in two K$_{0.3}$MoO$_3$ samples at $T=20$ K.} 
\label{spBB}
\end{figure}

\begin{figure}
\includegraphics[width=6.5cm]{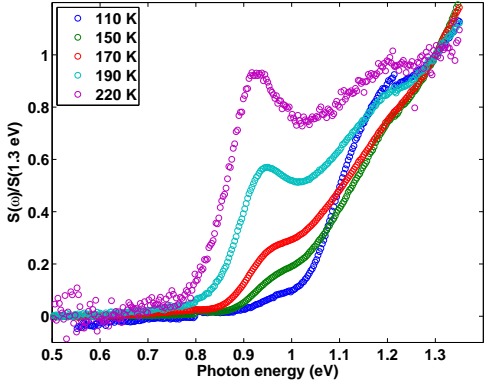}
\caption{Temperature evolution of normalized photoconduction in a NbS$_3$ sample with the most pronounced peak at $\hbar\omega = 0.9$~eV.} 
\label{spNbS3}
\end{figure}

Details of the energy structure can be obtained from spectral study of photoconduction. As it was noted above, strong non-monotonous temperature dependence of photoconduction (Figs. \ref{rtTaS3}, \ref{rtNbS3}) limits the temperature range available for study of spectral response in {\it o}-TaS$_3$, and especially for  K$_{0.3}$MoO$_3$.  Typical photoconduction energy spectra of {\it o}-TaS$_3$, K$_{0.3}$MoO$_3$ and NbS$_3$(I) are shown in Figs. \ref{spTaS$_3$}, \ref{spBB} and \ref{spNbS3}. All the spectra are normalized by the number of incident photons, $S(\omega) = \hbar\omega\delta G/W$. 
The most simple behavior is observed in K$_{0.3}$MoO$_3$ crystals (Fig. \ref{spBB}) where the gap edge is not complicated by any additional structure. 

In {\it o}-TaS$_3$ the spectrum could be separated into an intrinsic part and relatively narrow lines described by Gaussians \cite{naszzjetpl}. The amplitude of these spectral lines depends on the temperature, and at least for some of them  on the electric field \cite{comment}. Large amplitudes of the lines means huge amount of respective energy levels. It is very likely, therefore, that the lines observed in {\it o}-TaS$_3$ correspond to structural defects (e.g. to stacking faults in the crystal lattice or dislocations in CDW) rather than impurity levels. On our experience K$_{0.3}$MoO$_3$ is less subjected to formation of planar defects then {\it o}-TaS$_3$. In addition, the spectra of studied impure crystals are not much complicated by additional lines. This correlates with our experience that cleavage of doped crystals is much more complex in comparison with pure ones. Other words, doped crystals are more stable with respect to defect nucleations.

\begin{figure}
\includegraphics[width=6.5cm]{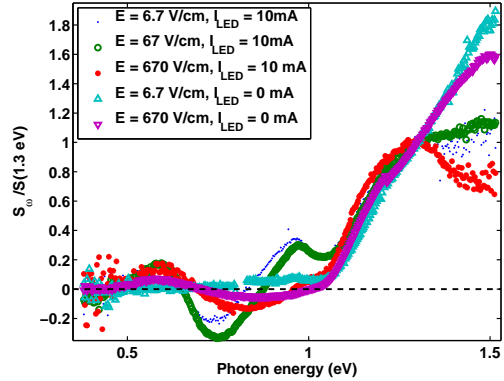}
\caption{Effect of additional illumination and high electric field on spectral dependencies of normalized photoconduction in NbS$_3$(I) sample at $T = 78$ K.} 
\label{spNbS3ET}
\end{figure}

Very interesting behavior is observed in NbS$_3$(I) where the photoconduction spectrum (including the mid-gap state found in this compound) can be substantially modified by application of the electric field, an additional illumination or both (Fig. \ref{spNbS3ET}) \cite{venera}.

\section{Impurity effect in {\it o}-TaS$_3$}
\label{impure}
It is well known that impurities in quasi-1D conductors lead to pinning of the CDW and suppress its giant dielectric constant. Do impurities also produce observable energy levels inside the Peierls gap and provide doping of CDW conductors? In addition, reduction of the transverse sizes of quasi-1D conductors leads to gradual appearance of features expected for 1D behavior) for nanoscale-sized samples (the power laws in temperature- and electric-filed dependent conduction \cite{1d}. Do moderate amount of impurities provides transition into a new state with impurity-stabilized Luttinger liquid  \cite{art}? In principle, as these questions could be answered on the basis of photoconduction study, it is worthwhile to undertake it for doped crystals of quasi-1D conductors.

\begin{figure}
\includegraphics[width=6.5cm]{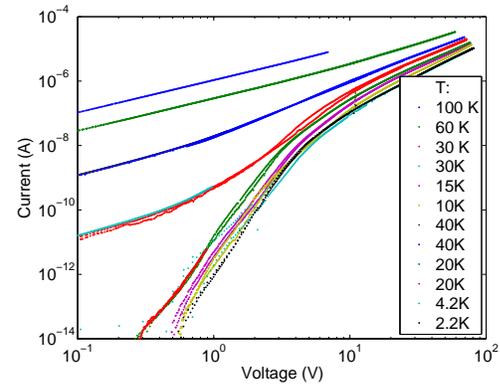}
\caption{Temperature set of I-V curves in Nb-doped {\it o}-TaS$_3$ sample.} 
\label{ivspain}
\end{figure}

We have undertaken such a study for {\it o}-TaS$_3$ with Nb impurities (impurity content 0.2 and 0.5 at.\%). Fig. \ref{ivspain} shows I-V curves of Nb-doped sample (0.5 at.\% of Nb) exhibiting non-linear I-V curves. Though the threshold field at $T>100$~K exceeds 100 V/cm, there is very strong low-temperature nonlinearity which can be described by two power laws, $I\propto V^\alpha$, with the exponents $\alpha =7$ and 2.3 (see, for comparison, Ref. \cite{1d}). 

\begin{figure}
\includegraphics[width=6.5cm]{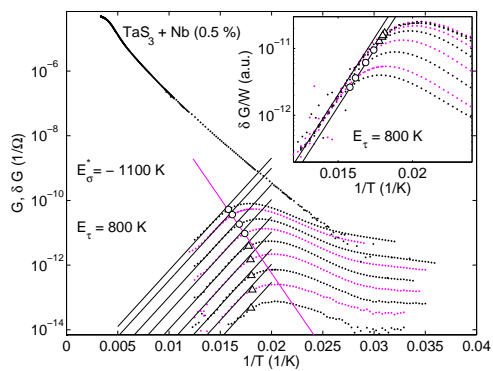}
\caption{Temperature dependences of conduction and photoconduction in Nb-doped {\it o}-TaS$_3$ sample. W=20, 10, 3, 1,  0.3, 0.1, 0.03, 0.01, 0.003 mW/cm$^2$.} 
\label{rdNP4}
\end{figure}

Fig. \ref{rdNP4} shows $G(T)$ and a set of photoconduction curves,  $\delta G(T,W)$, for {\it o}-TaS$_3$ with 0.5 at.\% of Nb impurities. The activation energy $E_\tau$ is systematically smaller and varies between 600 K and 800 K in tree Nb-doped samples, whereas $E_\sigma^*$ is roughly the same. $\tau_0$ is smaller in doped samples (Fig. \ref{tau}).
 Such a behavior may result from appearance of states inside the Peierls gap. 

\begin{figure}
\includegraphics[width=6.5cm]{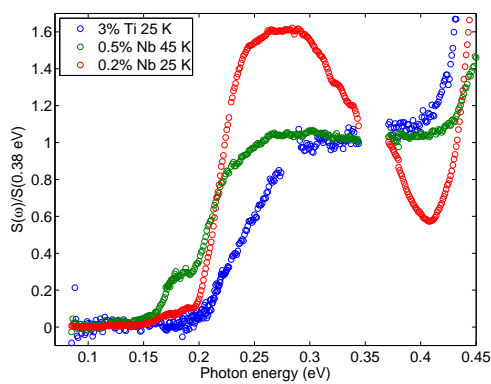}
\caption{Spectral dependencies of photoconduction in Ti- and Nb-doped {\it o}-TaS$_3$ samples.} 
\label{spspain}
\end{figure}

Fig. \ref{spspain} shows photo-current spectra of two Nb-doped samples and Ti-doped {\it o}-TaS$_3$ crystal (3 at. \% of Ti in the ampule). A sort of systematic spectra variation is seen from this figure: growth of impurity concentration suppresses the peak at $\hbar \omega\approx 0.25$ eV, and leads to appearance of intra-gap states.

\section{Conclusions}
Photoconduction study of quasi-1D conductors allows to distinguish the single-particle and collective {\it linear} conduction, investigate the effect of screening on collective transport and obtain interesting new details of the electronic energy structure of CDW conductors. A number of questions still remain to be open: the origin of low-temperature collective conduction in {\it o}-TaS$_3$ (including the shoulder in $G(T)$), the type of CDW gap (direct or indirect), identification of energy levels, the mechanism of quadratic recombination (radiative or phonon-assisted), the origin of low-temperature nonlinear conduction and voltage-enhanced photoconduction, {\em etc}. All these questions are definitely worthed further studying.
\\
\\
{\bf Acknowledgments}

We are grateful to R.E. Thorne (Cornell University), and F. Levy (Inst\' itute de Phys\' ique Appliqu\' ee, Lausanne) for supplying the crystals, and S.N. Artemenko, S. Brazovskii, N. Kirova,  and V.Ya. Pokrovskii for useful discussions. The support of RFBR and Department of Physical Sciences of RAS is acknowledged. The researches were performed in the framework of the Associated European Laboratory ``Physical properties of coherent electronic states in condensed matter'' between MCBT, Ne\' el Institute, C.N.R.S. and Kotel'nikov IRE RAS.




\begin{thebibliography}{00}
\bibitem{comment}Electric-field dependent peak is present in the sample with the ``shoulder'' in $G(T)$.
\end{thebibliography}
\end{document}